\documentstyle[pra,aps]{revtex}
\begin{document}
\draft
\twocolumn[\hsize\textwidth\columnwidth\hsize\csname 
@twocolumnfalse\endcsname                            
\title{Shifts and widths of collective excitations in
trapped Bose gases by the dielectric formalism}
\author{J\"urgen Reidl$^1$, Andr\'as Csord\'as$^2$, Robert Graham$^1$,
P\'eter Sz\'epfalusy$^3$}
\address{$^1${\it Fachbereich Physik, Universit\"at Gesamthochschule 
Essen,
45117 Essen, Germany}\\
$^2${\it Research Group for Statistical Physics of the
Hungarian Academy of Sciences,\\
P\'azm\'any P\'eter s\'et\'any 1/A, H-1117 Budapest, Hungary}\\
$^3${\it Department of Physics of Complex Systems,
E\"otv\"os University, P\'azm\'any P\'eter s\'et\'any 1/A,
H-1117 Budapest, Hungary,\\
and
Research Institute for Solid State Physics and Optics,
P.O. Box 49, H-1525 Budapest, Hungary}}
\date{\today}
\maketitle

\begin{abstract}
We present predictions for the temperature dependent shifts and
damping rates. They are obtained by applying the dielectric formalism
to a simple model of a trapped Bose gas.
Within the framework of the model
we use lowest order perturbation
theory to determine the first order correction to the results of
Hartree-Fock-Bogoliubov-Popov theory for the complex collective
excitation frequencies, and present numerical results for the 
temperature dependence of the damping rates and the frequency shifts.
Good agreement with the experimental values measured at JILA
are found for the $m=2$ mode, while we  find disagreements
in the shifts for $m=0$.
The latter point to the necessity of a non-perturbative
treatment  for an explanation of the temperature-dependence of
the m=0 shifts.
\end{abstract}

\pacs{03.75.Fi,05.30.Jp,67.40.Db}

\vskip2pc]                                            
\narrowtext                                           

\section{Introduction}

Since the discovery of Bose-Einstein condensation in traps a wealth of
experimental data on collective excitations
has appeared in the literature (for experimental
reviews see \cite{jila,mit}) waiting for theoretical explanation.
Some of the earliest measurements, still requiring a
firm theoretical analysis, were performed in oscillating
traps which permitted the selective excitation of different
excitation-modes \cite{Jin0,Jin,Mewes,kurn0}.
Common features of the above experiments are that the density of atoms
in the trap is relatively small and the temperature is extremely low.
Consequently, from the theoretical point of view the atoms can
be treated as a weakly interacting degenerate Bose gas, while the
interaction potential is well described by the $s$-wave approximation.
Due to the presence of the trap the whole
system is not translationally  invariant, and field equations of any
approximation must be solved in real-space, not in momentum space.
Furthermore the excitation spectrum is discrete rather than continuous
as in spatially homogeneous condensates like helium II.

Nevertheless, most theoretical approaches are based on the
natural generalization of one or the other of the homogeneous
descriptions to the inhomogeneous
case. The present paper also belongs to this line. It is based on the
dielectric formalism (see Ref. \cite{griffinbook} and further references
therein), first introduced for spatially homogeneous systems at zero
temperature \cite{ma,ksz0,wong},
later used at finite temperature \cite{szk,payne}, and recently 
generalized to inhomogeneous systems in \cite{bene}.
The great success of the dielectric
formalism lies in showing that the order parameter correlation function
(the one-particle Green's function) and the density-density correlation
function have the same spectra below the critical temperature.
In principle the dielectric formalism is valid at all temperatures
but still requires  to deal with infinitely many graphs to obtain exact
results. In practice however, one resorts to some approximations
for the proper and irreducible part of certain quantities of physical 
interest, which are the key quantities in the dielectric formalism. 
Following and extending
Ref. \cite{szk,bene} we shall base our investigation on a simple model 
for trapped weakly interacting Bose gases, which in
homogeneous systems (a) is valid at finite temperature, (b) satisfies
the generalized Ward identities, and (c) guarantees that both the Green's
function and the density-density autocorrelation spectra
exhibit the same excitations.
In addition the model we use, which builds on and extends a simpler one
discussed in detail in Ref. \cite{szk,griffinbook}, can be shown to be
related to an approximation used by Minguzzi and Tosi \cite{tosi}.

It is our main purpose here  to evaluate, within the approximation defined 
by our choice of the model,
the damping rates and frequency shifts and to compare the results with the
experimental
data \cite{Jin} of the JILA group. In principle we are also able
to calculate theoretical values for the MIT measurement, but
in our present approach we treat the fluctuations of the thermal
density perturbatively and consider only Landau damping.
This approach is justified if the condition $k_BT\gg \mu$ is fulfilled
which is only the case for the JILA measurements. Furthermore,
the high anisotropy of the MIT trap leads to
some numerical difficulties in the  code we use so far.
>From the theoretical side several papers have appeared in the
literature going  beyond the Hartree-Fock-Bogoliubov-Popov (HFBP)
theory \cite{HFBP}, which
is really necessary to take into account damping processes.
Most relevant papers beyond HFBP describe the damping process
\cite{liu,pita,giorgini,fedi0,fedi}
or  calculate the shifts by including
the anomalous average in such a way that the resulting
approximation is gapless \cite{hdb}. The approach applied in
Ref. \cite{fedi} is the second order Beliaev theory which is known to be
gapless. It treats both Beliaev and Landau dampings and also calculates
the shift of elementary excitations  in local density approximation along  
with their  damping.

In the present paper we wish to present a theory of the shifts and widths
of the {\it low-lying} modes. Our approach is
based on the dielectric formalism
 and achieves its
simplicity by a judicious selection of a subset
of graphs for the {\it proper part} of the physical quantities of interest.
The model  we shall investigate in detail in this work
accounts for the Landau damping, but cannot account for
Beliaev damping. This sacrifice
for gain in simplicity is not too big because
 in the temperature region where the
measurements we wish to understand are performed,
 and where we calculate the shifts and the
damping rates, the Beliaev damping of the excitations is negligible.

The paper is organized as follows. In Sec. \ref{sec:form} we briefly
summarize the general framework of the dielectric formalism for 
inhomogeneous systems and consider  an extended version of
the model approximation of Ref. \cite{szk,bene}.
The conditions for choosing the necessary
basic building blocks for the ladder approximation and
for the proper and irreducible quantities are spelled out. Then
 the quantities given in our formalism are related to the
fluctuations of the condensate and the
thermal density.
We show that neglecting the thermal density fluctuations
we recover
the usual Hartree-Fock-Bogoliubov-Popov equations \cite{HFBP}.
Then we derive the corrections of the
HFBP excitation energies  to
first order in the  thermal density fluctuations.
To solve the closed set of equations for the damping and the shifts still
requires some numerical work in the inhomogeneous case.
We briefly discuss our
numerical procedure   in Sec. \ref{sec:numerics}.
Sec. \ref{sec:results}. contains the  discussion of our results
 and the comparison with the experimental data
measured at JILA.
The last Section \ref{sec:conclusion}. is devoted to the
conclusions and to some final remarks.
 There we also compare
our model approximation with other approximations given in the 
literature, e.g. the already mentioned
treatment of all the  Beliaev diagrams by Fedichev et al.
\cite{fedi},
the kinetic equations of Ref. \cite{tosi} and the collisionless
Bolzmann equation \cite{bijlsma,stoof}.

\section{Formulation}
\label{sec:form}

Here we summarize the dielectric formalism first applied to 
inhomogeneous systems in \cite{bene}.
However, we shall not repeat the whole treatment, but rather 
concentrate on the key points, and indicate
the new further steps.

The Hamiltonian of our problem in second quantized form is
\begin{eqnarray}
\hat{H}&=&\int d^3r\, \hat\Psi^\dagger (\bbox{r})
\left(-{\hbar^2 \over 2m}\Delta +U(\bbox{r}) \right)
\hat\Psi (\bbox{r}) \nonumber\\
&+&{1 \over 2} \int d^3r_1 \int d^3r_2 \,
\hat\Psi^\dagger (\bbox{r}_1)\hat\Psi^\dagger (\bbox{r}_2)
v(\bbox{r}_1,\bbox{r}_2)\hat\Psi (\bbox{r}_1)
\hat\Psi (\bbox{r}_2), \nonumber\\
\label{eq:hamop}
\end{eqnarray}
where $\hat\Psi (\bbox{r})$ is the Bose field operator, $U(\bbox{r})$
is the trap potential, and $v(\bbox{r}_1,\bbox{r}_2)$ describes
the two-body interaction. In the following it is chosen as
\begin{equation}
v(\bbox{r},\bbox{r}')=g \delta(\bbox{r}-\bbox{r}') \equiv
{4 \pi \hbar^2 a \over m} \delta(\bbox{r}-\bbox{r}'),
\end{equation}
where $a$ is the $s$-wave scattering length and $m$ is the mass of the
atoms.
Throughout we shall restrict ourselves
to temperatures below the critical temperature. As usual, for $T<T_c$
we split off the condensate wave-function $\Phi_0(\bbox{r})$,
\begin{equation}
\hat\Psi (\bbox{r})= \Phi_0 (\bbox{r})+ \hat\Phi (\bbox{r}),
\end{equation}
where $\Phi_0(\bbox{r})= \langle \hat\Psi (\bbox{r})\rangle$ and
$\langle \ldots \rangle$ denotes thermal averaging
\begin{equation}
\langle \hat{A} \rangle = {\mbox{Tr} \hat{A}
e^{-\beta(\hat{H}-\mu\hat{N})}\over \mbox{Tr}
e^{-\beta(\hat{H}-\mu\hat{N})}}.
\end{equation}
Since we are interested in finite temperature excitations we use
Green's functions
\begin{equation}
G_{\alpha,\beta}(\bbox{r},\tau;\bbox{r}',\tau')=-
\langle T_\tau\left[ \hat\Phi_\alpha(\bbox{r},\tau)
\hat\Phi_\beta^\dagger(\bbox{r}',\tau')\right]\rangle
\label{eq:gfdef}
\end{equation}
with field operators in Matsubara representation $\hat\Phi_1 
(\bbox{r},\tau) \equiv \hat\Phi(\bbox{r},\tau)$ and 
$\hat\Phi_2 (\bbox{r},\tau) \equiv
\hat\Phi^\dagger (\bbox{r},\tau)$. 
The other key quantity we are interested  
in is the density autocorrelation function
\begin{equation}
\chi (\bbox{r},\tau;\bbox{r}',\tau') =-
\langle T_\tau \left[
\tilde{n}(\bbox{r},\tau)\tilde{n}(\bbox{r}',\tau')\right]\rangle,
\label{eq:chidef}
\end{equation}
where $\tilde{n}(\bbox{r})= \hat{n}(\bbox{r})-
\langle \hat{n}(\bbox{r})\rangle$. 
The finite temperature Green's functions 
(\ref{eq:gfdef}) and
the autocorrelation function (\ref{eq:chidef}) are functions of $\tau ,
\tau'$ via $\tau-\tau'$ only and are periodic with period $\beta\hbar$. 
Thus, one can
expand them as Fourier series. The Matsubara Fourier coefficients are
given by
\begin{eqnarray}
G_{\alpha,\beta}(\bbox{r},\bbox{r}',i\omega_n)&=&
\int_0^{\beta\hbar} d\tau \, e^{i\omega_n \tau}G_{\alpha,\beta}
(\bbox{r},\bbox{r}',\tau), \nonumber\\
\omega_n&=&{2n\pi \over \hbar\beta},
\end{eqnarray}
where $n$ is an integer. A corresponding expansion is made for $\chi
(\bbox{r},\tau;\bbox{r}',\tau')$ with coefficients
$\chi(\bbox{r},\bbox{r}',i\omega_n)$.
Retarded functions can be obtained by the usual analytic continuation
($i\omega_n \to \omega + i\eta$, $\eta$ infinitesimal).

It is useful to introduce the {\it proper part} of a quantity 
which is defined as
{\it the sum of diagrammatic contributions, which cannot be split into two
parts
by cutting a single interaction line}. In the following proper parts 
will be denoted by a tilde. By definition the density autocorrelation
function and its proper part fulfill
\begin{eqnarray}
\chi(\bbox{r},\bbox{r}',\omega)&=&\tilde\chi(\bbox{r},\bbox{r}',\omega)
\nonumber\\
&& +{1 \over \hbar}\int d^3r_1 \int d^3r_2 \,
\tilde\chi(\bbox{r},\bbox{r}_1,\omega) v(\bbox{r}_1,\bbox{r}_2)\nonumber
\\&&\times
\chi(\bbox{r}_2,\bbox{r}',\omega).
\end{eqnarray}
When $\chi(\bbox{r},\bbox{r}',\omega)$ has a pole in $\omega$, but its
proper part is nonsingular at the same $\omega$, then there exists an
eigenfunction $\xi (\bbox{r})$ satisfying
\begin{equation}
\xi (\bbox{r})={1\over \hbar} \int d^3r_1 \int d^3r_2 \,
\tilde\chi(\bbox{r},\bbox{r}_1,\omega) v(\bbox{r}_1,\bbox{r}_2)
\xi (\bbox{r}_2).
\label{eq:eigen}
\end{equation}
This eigenfunction can be identified with the total density fluctuation
$\delta n(\bbox{r})=\xi (\bbox{r})$ at the eigenfrequency given 
by the pole. In a similar way we get  the eigenfunctions
$\varphi_1,\varphi_2$ corresponding to
the 1-particle Green's functions $G_{\alpha,\beta}$
by solving the eigenproblem:
\begin{eqnarray}\nonumber
\varphi_\alpha(\bbox{r})&=&
\frac{1}{\hbar}\int d^3\bbox{r}_1\int
d^3\bbox{r}_2G^{HO}_{\alpha,\gamma}
(\bbox{r},\bbox{r}_1,
\omega)
\Sigma_{\gamma,\beta}(\bbox{r}_1,\bbox{r}_2,
\omega)\\
&&\times
\varphi_\beta(\bbox{r}_2)
\end{eqnarray}
(Here and in the following we adopt the convention that summation 
has to be taken over repeated
greek indices).
\begin{eqnarray}
G^{HO}_{11}(\bbox{r},\bbox{r}',\omega_n)&=&
G^{HO}_{22}(\bbox{r'},\bbox{r},-\omega_n)\nonumber\\
&=&\sum_j\frac{\displaystyle{\varphi_j^{HO}(\bbox{r})
\varphi_j^{HO}(\bbox{r}')^*}}{\displaystyle{\omega_n-\hbar^{-1}
(\epsilon_j^{HO}-\mu^{HO})}}
\nonumber\,\,,\\
G^{HO}_{12}(\bbox{r},\bbox{r}',\omega_n)&=&G^{HO}_{21}
(\bbox{r},\bbox{r}',\omega_n)=0
\nonumber
\end{eqnarray}
are the Green's
functions of the free harmonic
trap with the corresponding harmonic oscillator
excitation frequencies $\epsilon_j^{HO}$, eigenfunctions $\varphi_j^{HO}
(\bbox{r})$
and chemical potential $\mu^{HO}=(\hbar/2)(\omega_x+\omega_y+\omega_z)$,
where $\omega_{x,y,z}$ are the three trap frequencies.
$\Sigma_{\alpha,\beta}$ are the  selfenergies describing the
corrections due to the interactions.
The condensate density fluctuations
$\delta n_c(\bbox{r})$ given by perturbations of the condensate density
$n_c({\bbox r})=|\Phi_0({\bbox r})|^2$ due to 
excitations of single-quasiparticle modes
can be identified with
\begin{equation}
\delta n_c(\bbox{r})=\Phi_0(\bbox{r})\left(\varphi_1(\bbox{r})+
\varphi_2(\bbox{r})\right)
\label{eq:defdnc}\quad .
\end{equation}

Below $T_c$ the proper part  $\tilde\chi$ can be further decomposed
into {\it irreducibile} (also termed "regular") $\tilde\chi^{(r)}$ and {\it
reducible}
(also termed "singular") $\tilde\chi^{(s)}$ parts. We
call a diagram {\it irreducible} if {\it it cannot be split into two parts 
by cutting a single 
particle line}. The reducible part $\tilde\chi^{(s)}$ is related to
the existence
of the so-called anomalous proper vertex $\tilde\Lambda_\alpha$, which
contains all the proper diagrams with only one outer interaction line and
only one outer particle line, and which is due to the 
presence of the condensate below $T_c$:
\begin{eqnarray}
\tilde\chi(\bbox{r},\bbox{r}',\omega)&=&
\tilde\chi^{(r)}(\bbox{r},\bbox{r}',\omega)
+\tilde\chi^{(s)}(\bbox{r},\bbox{r}',\omega) \nonumber\\
\tilde\chi^{(s)}(\bbox{r},\bbox{r}',\omega)&=&
\int d^3r_1 \int d^3r_2 \, \tilde\Lambda_\alpha(\bbox{r},\bbox{r}_1,\omega)
\tilde{G}_{\alpha,\beta}(\bbox{r}_1,\bbox{r}_2,\omega)\nonumber\\
&&\times
\tilde\Lambda_\beta(\bbox{r}_2,\bbox{r}',\omega)\quad. \nonumber\\
\label{eq:denprop}
\end{eqnarray}
Here $\tilde{G}_{\alpha,\beta}(\bbox{r}_1,\bbox{r}_2,\omega)$ is the proper
part of the Green's function satisfying Dyson's equation with the proper 
part $\tilde{\Sigma}_{\alpha,\beta}(\bbox{r}_1,\bbox{r}_2,\omega)$ of the
selfenergy only. In Ref. \cite{bene} it is shown that the eigenfunctions
$\xi , \varphi_\alpha$ belonging to the
same eigenvalue $\omega$ are related by the
anomalous vertex $\Lambda_\alpha$
containing all the proper and improper vertex contributions
\begin{equation}
\xi(\bbox{r})=\int\,d^3\bbox{r}_1\,\Lambda_\alpha(\bbox{r},\bbox{r}_1,\omega)
\varphi_\alpha(\bbox{r}_1)\quad .
\label{eq:eigcon}
\end{equation}

So far no approximation has been made. Now we define approximate
expressions for the building blocks
$\tilde{\Lambda}^0_\alpha$ and $\tilde\chi^0$
of the ladder approximation in which we derive the proper quantities
$\tilde{\Lambda}_\alpha$ and $\tilde\chi^{(r)}$: \\
The trivial vertex function
\begin{equation}
\tilde{\Lambda}^0_1 (\bbox{r},\bbox{r}',\omega)=
\tilde{\Lambda}^0_2 (\bbox{r},\bbox{r}',\omega)=
\Phi_0(\bbox{r})\delta(\bbox{r}-\bbox{r}')
\label{eq:anomver0}
\end{equation}
and the bubble graph
\begin{eqnarray}
&&\tilde\chi^0(\bbox{r},\bbox{r}',\omega)=
\nonumber\\
&&=-{1 \over2 \beta\hbar}
\sum_n G^{HF}_{\alpha,\beta}(\bbox{r},\bbox{r}',i\omega_n)
G^{HF}_{\beta,\alpha}(\bbox{r}',\bbox{r},i\omega_n-\omega)
\nonumber\\
&&=-{1 \over \beta\hbar}
\sum_n G^{HF}_{1,1}(\bbox{r},\bbox{r}',i\omega_n)
G^{HF}_{1,1}(\bbox{r}',\bbox{r},i\omega_n-\omega)
\label{eq:chi0}
\end{eqnarray}
with the Green's function $G^{HF}_{\alpha,\beta}$ corresponding to  the
Hartree-Fock Hamiltonian
\begin{equation}
\hat{H}^{HF}(\bbox{r})
=-\frac{\hbar^2\nabla^2}{2m}+U(\bbox{r})
+2g n(\bbox{r}).
\end{equation}
The Green's functions $G^{HF}_{\alpha,\beta}$
satisfy the relations:
\begin{eqnarray}
(\hbar\omega-\hat{H}^{HF}(\bbox{r}))
G^{HF}_{11}(\bbox{r},\bbox{r}',\omega)&=&\hbar\delta(\bbox{r}-\bbox{r}')
\,\,,\\
G^{HF}_{22}(\bbox{r},\bbox{r}',\omega)&=&
G^{HF}_{11}(\bbox{r}',\bbox{r},-\omega)\,\,,\\
G^{HF}_{12}&=&G^{HF}_{21}=0\,\, .
\end{eqnarray}
We have already used  the Hartree-Fock Green's functions in
the bubble graph. We will see later that this approach turns out to be
consistent.\\
In the ladder approximation the proper contributions $
\tilde\chi^{(r)}(\bbox{r},\bbox{r}',\omega)$ are
derived by subsequent insertions of interaction lines into the
bubble diagrams. Therefore $
\tilde\chi^{(r)}(\bbox{r},\bbox{r}',\omega)$ is determined by the
selfconsistent equation:
\begin{eqnarray}
\tilde{\chi}^{(r)}(\bbox{r},\bbox{r}',\omega)&=&
\tilde{\chi}^0(\bbox{r},\bbox{r}',\omega)\nonumber\\
&&+\frac{g}{\hbar}\int\,d^3\bbox{r}_1\,
\tilde{\chi}^0(\bbox{r},\bbox{r}_1,\omega)
\tilde{\chi}^{(r)}(\bbox{r}_1,\bbox{r}',\omega)
\label{eq:chir}
\end{eqnarray}
For the proper vertex function $\tilde{\Lambda}_\alpha$
we take into account the
trivial vertex $\tilde{\Lambda}^0_\alpha({\bf r})$ and the first order
correction $\tilde{\Lambda}^{(1)}_\alpha (\bbox{r},\bbox{r}',\omega)=
g\tilde\chi^0(\bbox{r},\bbox{r}',\omega)\Phi_0(\bbox{r}')$.
In $\tilde{\Lambda}^{(1)}_\alpha $ we replace the single
interaction line by the T-matrix which defines the following equation for
the proper vertex function $\tilde{\Lambda}_\alpha$:
\begin{eqnarray}
\tilde{\Lambda}_\alpha(\bbox{r},\bbox{r}',\omega)&=&
\tilde{\Lambda}^0_\alpha(\bbox{r},\bbox{r}',\omega)\nonumber\\&&
+\frac{g}{\hbar}\int\,d^3\bbox{r}_1\,
\tilde\chi^0(\bbox{r},\bbox{r}_1,\omega)
\tilde{\Lambda}_\alpha(\bbox{r}_1,\bbox{r}',\omega)\,,
\label{eq:anomverr}
\end{eqnarray}
from which it can be determined selfconsistently.\\
The analogous Dyson equation for the
complete vertex function given by
$\Lambda_\alpha(\bbox{r},\bbox{r}',\omega)=\\
\tilde{\Lambda}_\alpha(\bbox{r},\bbox{r}',\omega)
+\frac{g}{\hbar}\int\,d^3\bbox{r}_1\,
\tilde\chi^{(r)}(\bbox{r},\bbox{r}_1,\omega)
\Lambda_\alpha(\bbox{r}_1,\bbox{r}',\omega)$
can be easily
expressed in terms of the building blocks of the ladder approximation:
\begin{eqnarray}
\Lambda_\alpha(\bbox{r},\bbox{r}',\omega)&=&
\tilde{\Lambda}^0_\alpha(\bbox{r},\bbox{r}',\omega)\nonumber\\
&&
+2\frac{g}{\hbar}\int\,d^3\bbox{r}_1\,
\tilde{\chi}^0(\bbox{r},\bbox{r}_1,\omega)
\Lambda_\alpha(\bbox{r}_1,\bbox{r}',\omega)
\label{eq:anomverc}
\end{eqnarray}
We identify the thermal density fluctuations
$\delta n_T(\bbox{r})=\delta n(\bbox{r})-\delta n_c(\bbox{r})$
by inserting (\ref{eq:anomverc}) and (\ref{eq:defdnc}) in
(\ref{eq:eigcon}) and using (\ref{eq:defdnc}) and (\ref{eq:anomver0}):
\begin{equation}
\delta n_T(\bbox{r},\omega)
=2\frac{g}{\hbar}\int\,d^3\bbox{r}_1\,\tilde{\chi}^0(\bbox{r},\bbox{r}_1,
\omega)\delta n(\bbox{r}_1,\omega)
\label{eq:dnt}
\end{equation}
Another way to derive this result is to consider the possible diagrams
in the linear response function $\chi_T$ for the thermal density where
$\chi_T$ is defined by:
\begin{equation}
\delta n_T(\bbox{r})=\int d^3{\bbox r}'\chi_T(\bbox{r},\bbox{r}',\omega)
\delta V(\bbox{r}')
\end{equation} and
$\delta V$ is an additional small perturbation
coupling to the density operator.\\
The only diagrams of our approximation for
$\chi$ not contributing to $\chi_T$ are
those which start with
the trivial vertex function $\tilde\Lambda_\alpha^0=\Phi_0$ on the
side coupling to the thermal density fluctuation:
\begin{eqnarray}
\nonumber
\chi_T(\bbox{r},\bbox{r}',\omega)&=&\int d^3\bbox{r}_1
\left[\tilde{\chi}(\bbox{r},\bbox{r}_1,\omega)
-\int d^3\bbox{r}_2\int d^3\bbox{r}_3\right.\\
&&\nonumber\left.
\tilde\Lambda^0_\alpha(\bbox{r},\bbox{r}_2,\omega)
\tilde{G}_{\alpha,\beta}
(\bbox{r}_2,\bbox{r}_3,\omega)
\tilde{\Lambda}_\beta(\bbox{r}_3,\bbox{r}_1,\omega)\right]
\nonumber\\&&
\times\left[\delta(\bbox{r}_1-\bbox{r}')
+\frac{g}{\hbar}\chi(\bbox{r}_1,\bbox{r}',\omega)\right]
\label{eq:chitv}\\
&=&
\int d^3\bbox{r}_1
\left[\tilde{\chi}^{(r)}(\bbox{r},\bbox{r}_1,\omega)\nonumber\right.\\&&
\left.
+ \frac{g}{\hbar}\int d^3\bbox{r}_2 
\tilde{\chi}^0(\bbox{r},\bbox{r}_2,\omega)
\tilde{\chi}^{(s)}(\bbox{r}_2,\bbox{r}_1,\omega)
\right]\nonumber
\\&&\times
\left[\delta(\bbox{r}_1-\bbox{r}')+\frac{g}{\hbar}
\chi(\bbox{r}_1,\bbox{r}',\omega)\right]
\end{eqnarray}
In the previous
step we have inserted the expression for $\tilde\Lambda^0_\alpha$
given by  equation (\ref{eq:anomverr}) into (\ref{eq:chitv}).
If we use $\delta V=\int \chi^{-1}\delta n$ we get
\begin{eqnarray}
\delta n_T(\bbox{r})&=&\int d^3\bbox{r}_1\int d^3\bbox{r}'
\chi_T(\bbox{r},\bbox{r}_1,\omega)
\chi^{-1}(\bbox{r}_1,\bbox{r}',\omega)\delta n(\bbox{r}')
\nonumber
\\
&=&\int d^3\bbox{r}_1\int d^3\bbox{r}_2\int d^3\bbox{r}'
\left[\tilde{\chi}^{(r)}(\bbox{r},\bbox{r}_1,\omega)
\right.\nonumber\\&&
\left.+\frac{g}{\hbar}\tilde{\chi}^0(\bbox{r},\bbox{r}_2,\omega)
\tilde{\chi}^{(s)}(\bbox{r}_2,\bbox{r}_1,\omega)
\right]\nonumber\\&&
\times\left[\chi^{-1}(\bbox{r}_1,\bbox{r}',\omega)
+\frac{g}{\hbar}\delta(\bbox{r}_1-\bbox{r}')\right]\delta n
(\bbox{r}')\end{eqnarray}
which reduces in the case of resonance ($\int\chi^{-1}\delta n=0$)
to the relation:
\begin{eqnarray}
\delta n_T(\bbox{r})&=&\int d^3\bbox{r}_1\int d^3\bbox{r}'
\left[\tilde{\chi}^{(r)}(\bbox{r},\bbox{r}_1,\omega)\right. \nonumber\\
&&\left.
+\frac{g}{\hbar}\tilde{\chi}^0(\bbox{r},\bbox{r}',\omega)
\tilde{\chi}^{(s)}(\bbox{r},\bbox{r}',\omega)
\right]\frac{g}{\hbar}\delta n
(\bbox{r}')
\label{eq:dnt2}
\end{eqnarray}
Using $\delta n=\delta n_c+\delta n_T$ the equation for
$\delta n_c$ has the form:
\begin{eqnarray}
\delta n_c(\bbox{r})&=&\int d^3\bbox{r}_1\int d^3\bbox{r}'
(\delta(\bbox{r}-\bbox{r}_1)-\frac{g}{\hbar}
\tilde{\chi}^0(\bbox{r},\bbox{r}_1,\omega))\nonumber\\&&
\times\tilde{\chi}^{(s)}(\bbox{r}_1,\bbox{r}',\omega)\frac{g}{\hbar}
\delta n(\bbox{r}')
\label{eq:dnc}
\end{eqnarray}
>From equations (\ref{eq:dnt2}) with (\ref{eq:chir}) and (\ref{eq:eigen})
it is straightforward  to derive  the final result
$\delta n_T=2(g/\hbar)\int \tilde{\chi}^0
\delta n$.

For the purpose of calculating $\delta n_c$ we need to
make additional approximations for the proper selfenergies
 $\tilde{\Sigma}_{\alpha,\beta}$.
We restrict ourselves to the proper and irreducible diagrams which
(a) are only 0-loop and 1-loop diagrams according to the
classification by Wong and Gould \cite{wong} (justified
by assuming $\delta n_T$ to be small and only calculating its first order
contributions),
(b) are connected with Landau damping 
(since we neglect the Beliaev damping) 
and  (c) do not contain anomalous Greensfunctions (Popov approximation).\\
First we introduce those selfenergies $\tilde{\Sigma}^0_{\alpha,\beta}$
appearing in the
Gross-Pitaevskii-equation for the condensate:
\begin{equation}
\tilde{H}_0\Phi_0(\bbox{r})=0.
\label{eq:gpeq}
\end{equation}
where $\tilde{H}_0$ is given by
\begin{equation}
\tilde{H}_0=-{\hbar^2 \over 2m}\Delta+U(\bbox{r})-\mu +
g|\Phi_0(\bbox{r})|^2
+2gn_T(\bbox{r}),
\label{eq:gpop}
\end{equation}
\begin{equation}
\tilde{\Sigma}^0_{\alpha,\beta}(\bbox{r},\bbox{r}',\omega)=\left(
g|\Phi_0(\bbox{r})|^2
+2gn_T(\bbox{r}_1)\right)
\left(\begin{array}{rl}
1&0\\
0&1
\end{array}\right)
\end{equation}
They contain the 0-loop diagram $ g|\Phi_0(\bbox{r})|^2$
together with the contributions from  the
stationary density of the non-condensed atoms:
\begin{equation}
n_T(\bbox{r})= \langle \hat\Phi^\dagger(\bbox{r}) \hat\Phi(\bbox{r})\rangle
\label{eq:nt}.
\end{equation}
This density can be calculated in two different ways: either in 
Popov approximation as in \cite{reidl}, which is a selfconsistent, gapless
approximation, but does not satisfy the Ward identities in the 
homogeneous case, or by using the Hartree-Fock approximation, which 
neglects the quasi-particle aspects of the thermal excitations but has
the great virtue of satisfying the Ward identities in the spatially
homogeneous case \cite{proc}. Therefore we choose this second method
in the following. However, it is remarkable that the final results for the
shifts and widths we obtain are very similar for both methods of 
calculation.
Due to conditions (a)-(c) we are left with only four  additional 
selfenergy
contributions which, using the same T-matrix approximation as before, 
can be written as:
\begin{equation}
\tilde{\Sigma}^{(1)}_{\alpha,\beta}(\bbox{r},\bbox{r}',\omega)=
\hbar\sigma(\bbox{r},\bbox{r}',\omega)
\left(\begin{array}{lr}
1&1\\
1&1
\end{array}\right)
\end{equation}
with
\begin{equation}
\sigma(\bbox{r},\bbox{r}',\omega)=
\frac{g^2}{\hbar^2}
\Phi_0(\bbox{r})\tilde{\chi}^{(r)}(\bbox{r},\bbox{r}',\omega)
\Phi_0(\bbox{r}')\,\,.
\end{equation}
The corresponding Greensfunctions
\begin{equation}
\tilde{G}^0_{1,2}=\tilde{G}^0_{2,1}=0,
\label{eq:gtoffd}
\end{equation}
\begin{equation}
\tilde{G}^0_{1,1}(\bbox{r},\bbox{r}',\omega)=
\tilde{G}^0_{2,2}(\bbox{r}',\bbox{r},-\omega)
\label{eq:gtdiag}
\end{equation}
satisfy
\begin{equation}
(\hbar\omega-\tilde{H}_0)\tilde{G}^0_{1,1}(\bbox{r},\bbox{r}',\omega)=
\hbar\delta(\bbox{r}-\bbox{r}'),
\label{eq:gteq}
\end{equation}
and $\sum_{\alpha\beta}\tilde{G}_{\alpha\beta}(\bbox{r},\bbox{r}',\omega)$
is determined by:
\begin{eqnarray}
\nonumber
\sum_{\alpha\beta}\tilde{G}_{\alpha\beta}(\bbox{r},\bbox{r}',\omega)&=&
\int d^3 \bbox{r}_1\int d^3 \bbox{r}_2
 \sum_{\kappa\lambda}
\tilde{G}^0_{\kappa\lambda}(\bbox{r},\bbox{r}_1,\omega)\\
&&\times
\big  
(\delta(\bbox{r}_1-\bbox{r}_2)\delta(\bbox{r}_2-\bbox{r}')
+\sigma(\bbox{r}_1,\bbox{r}_2,\omega)
\nonumber\\
&&\times
\sum_{\alpha\beta}\tilde{G}_{\alpha\beta}(\bbox{r}_2,\bbox{r}',\omega)\big)
\label{eq:sumdys}\,\,.
\end{eqnarray}
Inserting $\tilde{\chi}^{(s)}=\sum_{\alpha\beta}\tilde{\Lambda}_\alpha
\tilde{G}_{\alpha\beta}
\tilde{\Lambda}_\beta$ in equation (\ref{eq:dnc}) we get 
for $\delta n_c(\bbox{r})$:
\begin{eqnarray}
\delta n_c(\bbox{r})&=&\int d^3 \bbox{r}_1\int d^3 \bbox{r}_2
\int d^3 \bbox{r}_3\nonumber
\frac{g}{\hbar}\tilde\Lambda^0_\kappa(\bbox{r})
\tilde{G}^0_{\kappa\lambda}(\bbox{r},\bbox{r}_1,\omega)
\\&&
\times\left[\tilde{\Lambda}_\lambda
(\bbox{r}_1,\bbox{r}_2,\omega)\delta n(\bbox{r}_2)
+\frac{g}{\hbar}\tilde\Lambda^0_\lambda(\bbox{r}_1,\bbox{r}_3,\omega)
\right.\nonumber\\
&&\left.\times
\tilde\chi^{(r)}(\bbox{r}_3,\bbox{r}_2,\omega)
\delta n_c(\bbox{r}_2)
\right]
\label{eq:dnc2}
\end{eqnarray}
which may be rewritten as 
\begin{eqnarray}
\delta n_c(\bbox{r})&=&\int d^3 \bbox{r}_1\int d^3 \bbox{r}_2
\frac{g}{\hbar}\Phi_0(\bbox{r})
\sum_{\alpha\beta}\tilde{G}^0_{\alpha\beta}(\bbox{r},\bbox{r}_1,\omega)
\Phi_0(\bbox{r}_1)\nonumber\\&&
\times\displaystyle\left[\delta(\bbox{r}_1-\bbox{r}_2)\delta n(\bbox{r}_2)
+\frac{g}{\hbar}\tilde{\chi}^{(r)}(\bbox{r}_1,\bbox{r}_2,\omega)\right.
\nonumber\\&& \times \left.
(\delta n(\bbox{r}_2)+\delta n_c(\bbox{r}_2))
\displaystyle\right]\,\,.\label{eq:kingp}
\end{eqnarray}

To solve Eq. (\ref{eq:kingp})  is still very
difficult. Instead, we follow the procedure applied in \cite{bene}, where
the term $\frac{g}{\hbar}\tilde{\chi}^{(r)}$  
was treated as a small perturbation
and the excitation frequencies were determined using first order
perturbation theory. In the perturbation term 
$\frac{g}{\hbar}\tilde{\chi}^{
(r)}$ we may replace $\tilde{\chi}^{
(r)}$ by $\tilde{\chi}^{0}$ 
and $\delta n_c$ by $\delta n$ to first order. We then 
use eq.(\ref{eq:dnt}) to obtain
\begin{eqnarray}
\delta n_c(\bbox{r})&\approx&\int d^3 \bbox{r}_1
\frac{g}{\hbar}\Phi_0(\bbox{r})
\sum_{\alpha\beta}\tilde{G}^0_{\alpha\beta}(\bbox{r},\bbox{r}_1,\omega)
\Phi_0(\bbox{r}_1)
\nonumber\\
&&\times\left[\delta n_c(\bbox{r}_1)+2\delta n_T(\bbox{r}_1)\right]\,\,.
\label{eq:kingp2}
\end{eqnarray}

In the next step we show that the unperturbed problem
\begin{eqnarray}
\delta n_c^0(\bbox{r})&=&\frac{g}{\hbar}\int d^3 \bbox{r}_1\Phi_0(\bbox{r})
\sum_{\alpha\beta}\tilde{G}^0_{\alpha\beta}(\bbox{r},\bbox{r}_1,\omega)
\Phi_0(\bbox{r}_1)\nonumber\\
&&\times\delta n_c^0(\bbox{r}_1)
\end{eqnarray}
is equivalent to the
Hartree-Fock-Bogoliubov-Popov calculation of $\delta n_c^0$.
By dividing equation (\ref{eq:kingp2}) on both sides with
the condensate function $\Phi_0(\bbox{r})$ and afterwards
muliplying both sides from the left with $(\hbar\omega-\tilde{H}_0)
(-\hbar\omega-\tilde{H}_0)$ we derive an equation of the
form of the  diagonalized HFBP equations:
\begin{eqnarray}
\left[\tilde{H}_0^2(\bbox{r})-\hbar^2\omega^2\right]
\frac{\delta n_c(\bbox{r})}{\Phi_0(\bbox{r})}&=&
-2g\tilde{H}_0(\bbox{r})n_c(\bbox{r})
\nonumber\\&&\times\left(\frac{\delta n_c(\bbox{r})}
{\Phi_0(\bbox{r})}+2\frac{\delta n_T(\bbox{r})}
{\Phi_0(\bbox{r})}\right)
\label{eq:perturb}\,.
\end{eqnarray}
This equivalence can be shown by
performing the sum and the difference of 
the Hartree-Fock-Bogoliubov equations:
\begin{eqnarray}
&\left(
\begin{array}{cc}
\tilde{H}^0(\bbox{r})+gn_c(\bbox{r}) & gn_c(\bbox{r}) \\
g n_c(\bbox{r}) & \tilde{H}^0(\bbox{r})+ gn_c(\bbox{r})
\end{array}\right)&\left(
\begin{array}{c}
\varphi_1^{(j)}(\bbox{r})\\
\varphi_2^{(j)}(\bbox{r})
\end{array}\right)\nonumber\\
&=\hbar\omega_0^{(j)}\left(\begin{array}{c}
\varphi_1^{(j)}(\bbox{r})\\
-\varphi_2^{(j)}(\bbox{r})
\end{array}\right).&
\end{eqnarray}
We get the following diagonalized
equations for $\varphi_1^{(j)}\pm\varphi_2^{(j)}$:
\begin{eqnarray}
\nonumber
\hbar^2
\omega_0^{(j)}(\varphi_1^{(j)}(\bbox{r})+\varphi_2^{(j)}(\bbox{r}))&=&
\left[\tilde{H}_0^2(\bbox{r})+2g\tilde{H}_0(\bbox{r})n_c(\bbox{r})\right]\\
&&\times
(\varphi_1^{(j)}(\bbox{r})+\varphi_2^{(j)}(\bbox{r}))\\
\nonumber
\hbar^2
\omega_0^{(j)}(\varphi_1^{(j)}(\bbox{r})-\varphi_2^{(j)}(\bbox{r}))&=&
\left[\tilde{H}_0^2(\bbox{r})+2gn_c(\bbox{r})\tilde{H}_0(\bbox{r})\right]\\
&&\times(\varphi_1^{(j)}(\bbox{r})-\varphi_2^{(j)}(\bbox{r}))\,.
\end{eqnarray}
If we set
$\delta n_T=0$ (unperturbed case) in equation (\ref{eq:perturb})
we recover
the HFBP equation  where $\omega_0$ corresponds to  the eigenvalues
$\omega_0=\omega_0^{(j)}$ and $\delta n_c^0$ to the
corresponding eigenfunctions $\delta n_c^0=\xi_0^{(j)}=
\Phi_0(\varphi_1^{(j)}+\varphi_2^{(j)})$.\\
The first order correction due to the dynamics of the
thermal density   $\delta n_T$ (or in other words due to the
existence of $\tilde{\chi}^0$) can be calculated by
expanding
\begin{eqnarray}
\delta n_c&=&\delta n^0_c+\delta n_c^1 +\dots \\
\omega&=&\omega_0+\omega_1+\ldots,
\end{eqnarray}
where $\delta n_c^1$ and $\omega_1$ are the first order corrections.\\
Inserting this perturbation ansatz in
(\ref{eq:perturb}) the first order contributions must satisfy the equation:
\begin{eqnarray}
&&\left[\tilde{H}_0^2(\bbox{r})+2g\tilde{H}_0(\bbox{r})n_c(\bbox{r})
-\hbar^2\omega_0^2\right]\frac{\delta n_c^1(\bbox{r})}{\Phi_0(\bbox{r})}=
\nonumber\\&&
2\hbar^2\omega_1\omega_0\frac{\delta n_c^0(\bbox{r})}{\Phi_0(\bbox{r})}
-4g\tilde{H}_0(\bbox{r})n_c(\bbox{r})\frac{\delta n_T(\bbox{r})}
{\Phi_0(\bbox{r})}\,\,.
\end{eqnarray}
Multiplying this equation from the left with
$(\varphi_1(\bbox{r})-\varphi_2(\bbox{r}))$ and integrating over
space the left side of the equation vanishes. Due to
the normalization condition  $\int d^3\bbox{r}
(\varphi_1(\bbox{r})-\varphi_2(\bbox{r}))(\varphi_1(\bbox{r})+
\varphi_2(\bbox{r}))=1$ and
the relation  $ \tilde{H}_0(\varphi_1-\varphi_2)=
\hbar\omega_0(\varphi_1+\varphi_2)$
we get the following result for $\omega_1$:
\begin{eqnarray}
&&
\hbar\omega_1=2g  \int d^3r \,
\delta n_c^*(\bbox{r}) \delta n_T(\bbox{r})\\
&&={4g^2 \over \hbar}  \int d^3r \int d^3r' \,
\delta n_c^*(\bbox{r}) \tilde{\chi}^{0}(\bbox{r},\bbox{r}',\omega_0) 
\delta n_c(\bbox{r}').
\label{eq:firstcorr}
\end{eqnarray}

This is one of the basic results used in the following  for the calculation 
of shifts and widths of collective excitations. 
The consistency condition
for the applicability of perturbation theory is that 
$\omega_0\gg |\omega_1|$
must hold, which we shall check below.

Next, let us concentrate on the quantity
$\tilde\chi^0(\bbox{r},\bbox{r}',\omega_0)$.
If one expresses $G^{HF}_{1,1}(\bbox{r},\bbox{r}',i\omega_n)$
in terms of the eigenvalues and eigenvectors of the Hartree-Fock operator
$H^{HF}$:
($H^{HF}\varphi_i^{HF}=\varepsilon_i^{HF} \varphi_i^{HF}$)
\begin{equation}
G^{HF}_{1,1}(\bbox{r},\bbox{r}',i\omega_n)= \sum_i
{\varphi^{HF}_i (\bbox{r}) \varphi^{HF*}_i (\bbox{r}')\over
i\omega_n-\varepsilon^{HF}_i/\hbar }\,\,,
\end{equation}
then the standard Matsubara sum in Eq. (\ref{eq:chi0}) can be easily 
performed
\begin{eqnarray}
\tilde\chi^0(\bbox{r},\bbox{r}',i\omega_n)&=& \nonumber
\sum_{i,j} \left(n(\varepsilon^{HF}_j)-n(\varepsilon^{HF}_i) \right)\\
&&\times
{
\varphi_i^{HF} (\bbox{r}) \varphi^{HF*}_i (\bbox{r}')
\varphi_j^{HF} (\bbox{r}') \varphi^{HF*}_j (\bbox{r})\over i\omega_n
-(\varepsilon^{HF}_i- \varepsilon^{HF}_j)/\hbar},
\end{eqnarray}
where $n(\varepsilon^{HF}_i)$ is the Bose-factor.
>From the form of the denominator it is clear that we treat processes
leading to Landau damping, but not to Beliaev damping.
In calculating
the retarded function $\tilde\chi^0$ via the analytic continuation
$i\omega_n \to \omega_0 +i\eta$ its imaginary part contains a series
of Dirac-delta peaks due to the discrete spectrum of $H^{HF}$.
In order to get finite damping (i.e., finite imaginary part of $\omega_1$
in Eq. (\ref{eq:firstcorr})), which was measured in experiments,
one needs some smoothing over the Dirac-delta peaks.
In principle, for the shift there is no need for smoothing, but one
must use the same procedure both for the real and the imaginary part
of $\tilde\chi^0$, otherwise they do not fulfill the
Kramers-Kronig relations.
The smoothing is done by using the semiclassical or 
local density approximation.
First, we calculate the semiclassical Green's function. To do this,
let us consider the representation of the Green's function
in the center-of-mass, $\bbox{k}$ representation, defined as:
\begin{equation}
G^{HF}_{1,1}(\bbox{r},\bbox{r}',\omega)= \int
{d^3k \over (2\pi)^3} e^{i\bbox{k}(\bbox{r}-\bbox{r}')}
G^{HF}_{1,1}((\bbox{r}+\bbox{r}')/2,\bbox{k},\omega)
\label{eq:gtmix}
\end{equation}
For  this representation the semiclassical Green's function
is simply
\begin{equation}
G^{HF(sc)}_{1,1} (\bbox{R},\bbox{k},i\omega_n)=
{\hbar \over i\hbar\omega_n -(\epsilon_{\bbox{k}}-\mu(\bbox{R}))},
\label{eq:scgf}
\end{equation}
where
\begin{equation}
\epsilon_{\bbox{k}}={\hbar^2 k^2 \over 2m}, \quad \mu(\bbox{R})=
\mu-U(\bbox{R})-2gn_c(\bbox{R})-2gn_T(\bbox{R}).
\end{equation}

It is easy to show that if one uses the mixed representation for
$\tilde\chi^0$, similarly defined as in (\ref{eq:gtmix})
\begin{equation}
\tilde\chi^0(\bbox{r},\bbox{r}',\omega)=\int {d^3k \over (2\pi)^3}
e^{i\bbox{k}(\bbox{r}-\bbox{r'})} \tilde\chi^0
\left((\bbox{r}+\bbox{r}')/2,\bbox{k} ,\omega \right)
\label{eq:chimix}
\end{equation}
then in the new representation Eq.(\ref{eq:chi0}) takes the form
\begin{eqnarray}
\tilde\chi^0(\bbox{R},\bbox{k},i\omega_n)&=&
-{1 \over \beta\hbar} \sum_p \int {d^3k' \over (2\pi)^3}
G^{HF}_{1,1}(\bbox{R},\bbox{k}+\bbox{k}',i\omega_p)\nonumber\\
&&\times
G^{HF}_{1,1}(\bbox{R},\bbox{k}',i\omega_p-i\omega_n).
\nonumber\\
\end{eqnarray}
Performing the Matsubara sum, but now with the semiclassical 
Green's function
(\ref{eq:scgf}), after the analytic continuation we arrive at
\begin{equation}
\tilde\chi^0(\bbox{R},\bbox{k},\omega)=
\int {d^3k' \over (2\pi)^3}
{n_{\bbox{k}'} -n_{\bbox{k}+\bbox{k}'} \over
\omega-(\epsilon_{\bbox{k}+\bbox{k}'} -\epsilon_{\bbox{k}'})/\hbar +i\eta},
\label{eq:chitlda}
\end{equation}
where $n_{\bbox{k}}$ is the Bose-factor appearing in the
semiclassical (or local-density) approximation:
\begin{equation}
n_{\bbox{k}}={1 \over
\displaystyle{e^{\beta\left({\hbar^2 k^2 \over 2m} 
-\mu(\bbox{R})\right)}-1}}.
\end{equation}
In the region where the condensate density vanishes the condition
$\mu(\bbox{R})\approx0$ holds. As a result, in our semiclassical 
approximation we would there get a logarithmic singularity
of $\tilde\chi^0(\bbox{R},\bbox{k},w)$
at  $k=|\bbox{k}|=\sqrt{2m\hbar\omega}$
leading to large contributions from frequencies around
zero, which are completely artificial, since the trapped system 
has no levels below a certain lowest-lying level.
We therefore suppress these unphysical logarithmically singular 
contributions by
choosing the lowest-lying energy level $\omega_{min}=2.12\omega_\rho$
of the Hartree-Fock operator
as a natural  lower energy cut-off in our application of the Hartree-Fock 
and  local density approximation.
Here $\omega_\rho$ is the smallest trap frequency.
We thus ensure that  all the contributions to frequency shifts and damping
rates come only from interactions with modes in
the physical energy region above this lowest Hartree-Fock energy level.
In Fig. \ref{fig:dNode} we plot the spectral contributions
to the thermally excited atoms as a function of the frequency for both,
the local density
approximation and the histogram of energy-levels obtained by the
direct diagonalization of the Hartree-Fock Hamiltonian.
The agreement above our cut-off proves to be rather good so that we
can expect to get reliable results in the
local density approximation.
With this cut-off the imaginary part of $\tilde\chi^0$ becomes:
\begin{eqnarray}&&
\Im \tilde\chi^{(r)}(\bbox{R},\bbox{k},\omega)=
{m^2 \over 4\pi \hbar^3 k \beta} \nonumber\times\\&&
\ln
\left[ {1-\exp \left( -\beta \left(\frac{
(\hbar \omega -\epsilon_{\bbox{k}})^2}{
4\epsilon_{\bbox{k}}} -\mu(\bbox{R})\right) \right) \over
 1-\exp \left( -\beta \left(\frac{
(\hbar \omega +\epsilon_{\bbox{k}})^2}{
4\epsilon_{\bbox{k}}} -\mu(\bbox{R})\right) \right)
}\right]
\end{eqnarray}
for $(\frac{\omega m}{\hbar k}-\frac{k}{2})^2\geq
\max\left[2m\hbar^{-2} (\mu(\bbox{R})+\hbar\omega_{min}),0\right]$.
In the opposite case we have due to the cut-off
the following expression:
\begin{eqnarray}
\Im \tilde\chi^{(r)}(\bbox{R},\bbox{k},\omega)&=&
{m^2 \over 4\pi \hbar^3 k \beta}\nonumber\\&& 
\times\ln
\left[ {1-\exp \left( -\beta \hbar\omega_{min}\right) \over
 1-\exp \left( -\beta
\hbar \left(\omega_{min}+\omega\right)\right)
}\right]
\end{eqnarray}
for $(\frac{\omega m}{\hbar k}-\frac{k}{2})^2\le
\max\left[2m\hbar^{-2} (\mu(\bbox{R})+\hbar\omega_{min}),0\right]$.
After performing the principal value integral the real part can be
written as
\begin{eqnarray}
\Re \tilde\chi^{(r)}(\bbox{R},\bbox{k},\omega)&=&
{m \over 4\pi^2 \hbar q} \int_{\hbar\sqrt{2m\epsilon(\bbox{
R})}} k'\, dk' \, n_{\bbox{k}}\nonumber\\ &&\times
\ln\left[ {a(k,k',\omega,\bbox{R})
b(k,k',\omega,\bbox{R})\over c(k,k',\omega)d(k,k',\omega)}\right]
\end{eqnarray}
with
\begin{eqnarray}\nonumber
&&a(k,k',\omega,\bbox{R})=\max\left[\frac{k^2}{2m}-\frac{kk'}{m},
\frac{\epsilon(\bbox{R})}{\hbar^2}-\frac{k'^2}{2m}\right]-\hbar^{-1}
\omega\\&&
\nonumber
b(k,k',\omega,\bbox{R})=\min\left[\frac{kk'}{m}-\frac{k^2}{2m},
\frac{k'^2}{2m}-\frac{\epsilon(\bbox{R})}{\hbar^2}\right]-\hbar^{-1}
\omega\\&&
\nonumber
c(k,k',\omega)=\frac{k^2}{2m}+\frac{kk'}{m}-\hbar^{-1}\omega\\&&
\nonumber
d(k,k',\omega)=-\frac{k^2}{2m}-\frac{kk'}{m}-\hbar^{-1}\omega
\end{eqnarray}
$\epsilon(\bbox{R})$ is an abbreviation for\\
$\epsilon(\bbox{R})=
\max\left[\mu(\bbox{R})+\hbar\omega_{min},0\right]$.

\section{Numerical methods}
\label{sec:numerics}

In our perturbative calculation of $\omega_1$ we should solve
first the unperturbed problem for $\omega_0$ and $\delta n_c(\bbox{r})$.
This step requires to solve the Gross-Pitaevskii (GP) and the HFBP
equations.
We already have reported on our selfconsistent algorithm in \cite{reidl}
for this problem, which solves the GPE for $\Phi_0(\bbox{r})$ and for the
chemical potential using multigrid methods, while the HFBP equations 
are solved in  
a large
truncated basis. As a result, we have all the necessary inputs,
namely $\Phi_0(\bbox{r})$, $\varphi_1(\bbox{r})+\varphi_2(\bbox{r})$, 
$\omega_0$,
$n_T(\bbox{r})$ for performing
the second step: the use of Eq. (\ref{eq:firstcorr}) in connection with
the retarded density auto-correlation function Eqs. (\ref{eq:chimix}) and
(\ref{eq:chitlda}).

This step is straightforward, but from the numerical
point of view it requires to cope with six-dimensional space-integrals
and three-dimensional momentum-integrals. To reduce the number of numerical
integrations we proceed as follows.
If  we introduce $C$ as an intermediate quantity for
\begin{equation}
C(\bbox{R},\bbox{k})=\int d^3r\, e^{-i\bbox{k}\bbox{r}}
\delta n_c^*(\bbox{R}-{\bbox{r}\over 2}) 
\delta n_c(\bbox{R}+{\bbox{r}\over 2})
\end{equation}
then Eq. (\ref{eq:firstcorr}) can be written as
\begin{equation}
\hbar\omega_1= 4{g^2 \over\hbar} \int d^3R \int {d^3k \over (2\pi)^3} \,
\tilde\chi^0(\bbox{R},-\bbox{k},\omega_0)C(\bbox{R},\bbox{k}).
\label{eq:firstcorr1}
\end{equation}
Next, let us expand $\delta n_c(\bbox{r})=\Phi_0(\bbox{r})
(\varphi_1(\bbox{r})+
\varphi_2(\bbox{r}))$
in some {\em isotropic} harmonic oscillator basis as
\begin{equation}
\delta n_c(\bbox{r})=\Phi_0(\bbox{r})(\varphi_1(\bbox{r})+
\varphi_2(\bbox{r}))=
\sum A_i f_i(\bbox{r}),
\end{equation}
where the set of basis functions $f_i (\bbox{r})$ have
the same angular momentum quantum number and
parity with respect to reflection on the $x$-$y$ plane as the unperturbed
elementary excitation in question. The next task  is to find a connection
formula, which transforms any product $f_i^*(\bbox{r}) f_j (\bbox{r}')$
to the mixed $(\bbox{R},\bbox{k})$ representation. This step requires
some straighforward but rather lengthy calculation, but the advantage we
gain is that due to the facts  that 
$\tilde\chi^{(r)}(\bbox{R},\bbox{k},\omega_0)$
is axially symmetric in $\bbox{R}$ for
axially symmetric harmonic oscillator trap potential
and isotropic in $\bbox{k}$
we can perform the integrations
over the azimuthal angle of $\bbox{R}$ and over the angles of $\bbox{k}$  in
(\ref{eq:firstcorr1}). At the end, apart from discrete sums, we have two
spatial integrals in the
radial and in the $z$ directions of $\bbox{R}$ and one momentum integral
for the imaginary part  and two subsequent momentum integrals
for the real part of $\omega_1$. Of course, the remaining integrals
must be performed numerically.

The form of the above mentioned connection formula which we
actually used is based on the following facts:
First, the product $f_i^*(\bbox{r}) f_j (\bbox{r}')$ is a solution of the
two-body, non-interacting  harmonic oscillator problem, which is also 
separable in center-of-mass and relative coordinates. 
Thus,  expanding it  in the
other basis containing products of harmonic oscillator
wave-functions depending on
$\bbox{R}=(\bbox{r}+\bbox{r}')/2$ and $\bbox{r}-\bbox{r}'$  the expansion
will contain only a finite number of new basis functions from
the energy shell $\varepsilon_i+\varepsilon_j$. Secondly, 
the Fourier-transform of the harmonic oscillator wave-functions 
(which we must perform with respect
to the relative coordinates) are basically the same as in real space.

\section{Results}
\label{sec:results}
We present our numerical results in  a way which is directly comparable
with the measurement \cite{Jin}. In Figs. \ref{fig1} and \ref{fig2}
we put on the horizontal axes $T'$, which is proportional to the RF
frequency used in evaporative cooling and is also the same quantity,
 used in Fig. 1 of Ref. \cite{Jin}. From that figure we take the same
$T=T(T')$ and $N_0=N_0(T')$, which were determined experimentally.
Also the same anisotropy $\omega_z/\omega_0$ as in the experiment was used.
We  present our results as a function of the experimental $T'$
because different experimental points belong to different physical
conditions since neither $N$ nor $N_0$ could be kept fixed in  the 
experiment.

In Fig. \ref{fig1} we plot the imaginary part of $\omega_1$ together
with the measured damping rates for the $m=0,2$ modes.
For temperatures $T>0.6 T_c$ where the Landau damping arising from the 
bubble graph $\tilde\chi^0(\bbox{r},\bbox{r}',\omega)$ is the dominant
process our results are inside the error bars. At lower temperatures
the calculated damping rates are slightly above the measured ones in 
agreement with the prediction of calculations for the homogeneous system.
There the values for the damping rates due to our model calculation 
are larger with a factor of $\frac{32}{9\pi}$ compared with the 
treatment of all Beliaev graphs to second order.
This is not surprising, because the neglection of the Beliaev damping
and the use of local density approximation is only justified for temperatures
satisfying the condition $k_B T\gg \mu$, only valid for $T\gg 0.6 T_c$. The 
same condition is necessary to neglect the quasi-particle character of the 
additional  excited particles contributing to the selfenergies 
via scattering 
processes.
Comparing our results for the damping rates with the values of the
unperturbed $\omega_0$ it is manifest that $\Im \omega_1 \ll \omega_0$. The
real part of $\omega_1$ is also much smaller then $\omega_0$ for
all the calculated points. Therefore the criterium for
the applicability of  perturbation  theory is justified.

In Fig. \ref{fig2} we plot $\omega_0+\Re \omega_1$ together with
the experimentally measured frequencies. For $m=2$ we have
good agreement for temperatures above $T\ge0.6 T_c$ and slightly larger 
negative shifts compared to the experimental values below
$T=.6T_c$. The results for the $m=0$ mode show the 
same slight difference in the low  temperature region. 
But approaching higher temperatures the experimental and the numerical results
predict a different behavior for the excitation energies. 
While experimentally they are increasing we found decreasing values 
in our model calculation.  The
same qualitative discrepancy has also been found by
Hutchinson et al. in \cite{hdb} using a different approach to the
problem. They also reported good agreement with the measured
frequencies for $m=2$, but disagreement for $m=0$.
A possible explanation for the above discrepancies has been given by
Bijlsma and Stoof in Ref. \cite{bijlsma} who used the quantum
Boltzmann-equation in the collisionless regime and calculated the
low-lying excitation frequencies using a variational approach for
the $m=0,2$ modes. They found that
there are two nearby $m=0$ modes, one of them describing the in-phase the
other one the out-of-phase oscillations of the condensate and the thermal
cloud. They concluded  that it might be possible that in the experiment
both are excited together and that in Ref. \cite{Jin} only the upper one
or a superposition of both might be plotted.
If there are two nearby $m=0$ modes we must explain why in our
calculation we have
obtained only
one of them (and furthermore not the experimentally measured one).
It is clear that
by our perturbative treatment we
fixed ourselves
to one of them by calculating the corrections to
the Hartree-Fock-Bogoliubov-Popov equations for
$\delta n_T\ll\delta n_c$ and we followed it in
changing the temperature. Our results correspond to the  branch describing
the out-of-phase motion calculated in Ref. \cite{bijlsma,stoof}.
To see this we recall our formula  for
the corrections $\omega_1=2g  \int d^3r \,
\delta n_c^*(\bbox{r}) \delta n_T(\bbox{r})$
and the fact that $\omega_1$ is found to be negative in our results.
Therefore the spacial average over the amplitude product of
$\delta n_T$ and $\delta n_c$ is negative which means that $\delta n_T$ and
$\delta n_c$ are out-of-phase.\\
The in-phase  mode can accordingly be obtained, if, in fact, there
is  another nearby $m=0$ mode in our model approximation, if we do not
use perturbation theory for solving (\ref{eq:perturb}). Instead we should
consider
the eigenvalues of the bubble graph $\tilde\chi^0$ corresponding to
the thermal density fluctuation $\delta n_T$ and calculate their
perturbation due to $\delta n_c$.
This possibility
is supported by the early study of the behavior of the poles 
of the density-density
autocorrelation function in the homogeneous Bose gas \cite{szk}, where
it was shown that there are more, but purely damped modes. They remain purely  
overdamped above $T_c$ in accord with a recent experiment \cite{Ketterle}.
It can be shown that one of the overdamped damped modes corresponds
to $\delta n_T\gg\delta n_c$
and that it is a mode where $\delta n_T$ and $\delta n_c$
relax in phase \cite{fliesser}. In addition it is clear that in
a harmonic trapping potential these damped modes will
acquire a nonzero real
part of their oscillation frequency.
However to really demonstrate this for the  eigensolutions of
the density-density
autocorrelation function in the inhomogeneous case,
and furthermore to show that in this case the missing mode
can be described in this way requires further analytical and
numerical studies.

\section{Conclusions and further remarks}
\label{sec:conclusion}

In this paper
we have investigated
the predictions of the dielectric
formalism applied to the model of a Bose gas obtained by
extending simpler models discussed in the literature
\cite{griffinbook,szk,bene}.
The proper and irreducible building blocks
$\tilde\chi^{(r)}$ and $\tilde\Lambda$ are already the
result of summing up higher order contributions in the form of a ladder
approximation  only summing up  geometric series of the
bubble graph $\tilde\chi^0$.
>From the second order Beliaev diagrams we keep
for the irreducible selfenergies only those which
are supposed to give large contributions.
Since Landau damping dominates Beliaev damping in the temperature
region $T\gg0.6T_c$ of the measurement at JILA we neglect the diagrams
connected with Beliaev damping.
Following the classification of Wong and Gould \cite{wong} 
we reduce the amount of
diagrams further to only  0-loop and 1-loop diagrams.
In addition we neglect all the selfenergy diagrams containing
anomalous Green's functions. To find
poles of the density-density correlation function, i.e.,  collective
excitations, we used first order perturbation theory, where the proper
and regular part of the density-density correlation function played
the role of the perturbing operator. We showed that the physical content
of the unperturbed problem is equivalent to find elementary excitations
in the Hartree-Fock-Bogoliubov-Popov approximation.
The fact that all the
first order corrections to the Hartree-Fock-Bogoliubov-Popov equations
can be described with one  bubble graph $\tilde\chi^0$ permits
to draw conclusions about the behavior of the quantities $\delta n_c$ and
$\delta n_T$ from   our
numerical results. Due to the negative energy shifts obtained in
our numerical results
we can identify their motion to be out-of-phase.
Since we establish our model on the basis of the dielectric formalism
we achieve the required correspondence of the spectra of
the Green's functions and the density auto-correlation function,
ensuring  gapless spectra of quasi-particle modes and density modes
in the homogeneous limit.
The approximate version (\ref{eq:kingp2}) of our equation  (\ref{eq:kingp})
agrees with
the dynamic Hartree-Fock-Bogoliubov-Popov theory presented
by Minguzzi and Tosi \cite{tosi}.
They derive (\ref{eq:kingp2}) by linearizing the time-dependent
Gross-Pitaevskii equation around its stationary solution $n(\bbox{r},t)=
n_c(\bbox{r})+\delta n_c(\bbox{r},t)$ and  (\ref{eq:dnt}) by an
equivalent linearization of the Hartree-Fock equation for the
non-condensate field operator.
They show that the correct  bubble graph $\tilde\chi^0$
connected with (\ref{eq:dnt}) is a product of Green's functions in
Hartree-Fock approximation. \\
Bijlsma and Stoof \cite{bijlsma} also  consider the  equation
(\ref{eq:dnt}) and use
the dispersion relation of the Hartree-Fock operator in
local density approximation for their calculations.
In contrast to us they examine the collisionless Bolzmann equation
for the Wigner function and solve the equations with a
variational approach, which cannot describe any damping of the excitations.
Whereas the local density approximation we and \cite{bijlsma} use is better
for large
condensates the approach of Fedichev et al. \cite{fedi0}
is better suited for
the opposite case of small condensates due to the assumption
of hard chaos, which can only be fulfilled for small condensates.
In this approach the diagrams are evaluated by integrating along
the classical trajectories.

 We have found good agreement
both for the shift and the damping with the experiment for $m=2$,
agreement also for the damping-rate of the $m=0$ mode,
 but disagreement similar to \cite{hdb} in the frequency shift for the
 $m=0$ mode.
We explained this discrepancy with observation
by an artefact of our perturbative treatment and by applying the
argument of Ref. \cite{bijlsma,stoof}, namely by supposing an other
nearby lying $m=0$ mode.

To decide, whether a nonperturbative treatment of the problem, but
still within the framework of our model approximation, can show
the existence of the missing $m=0$ mode requires an extended numerical
study. Our future plan is to find the poles in a nonperturbative way and
study the decoupling of the single particle and  the collective excitations
for  $T \to T_c-0$, along the lines of  the calculation for homogeneous
system of Ref. \cite{szk}.
A further natural extension of this work could be to study further
and more difficult, but numerically still manageable model approximations.

\acknowledgements

This work has been supported by a project of the Hungarian Academy
of Sciences and the Deutsche Forschungsgemeinschaft under Grant No. 95.
R. G. and J. R. wish to acknowledge support by the Deutsche
Forschungsgemeinschaft through the Sonderforschungbereich 237
"Unordnung and gro\ss e Fluktuationen". Two of us (A. Cs. and P. Sz.)
would like to acknowledge support by the Hungarian National Scientific
Research Foundation under Grant Nos. OTKA T017493, T025866 and F020094,
and by the Ministry of Education of Hungary under Grant No. FKFP1059/1997.

\begin{figure}
\caption{Spectral contributions to the thermally excited atoms
$\frac{dN}{d\omega}(\omega)$ as a function of the frequency $\omega$.
We compare the result of the
local density approximation (open circles) with the histograms
from the
direct diagonalization for $T=95$ nK  and $N_c=7500$.}
\label{fig:dNode}
\end{figure}
\begin{figure}
\caption{Damping rates of the $m=0$ (triangles) and $m=2$
(circles) modes as a function of  $T'$ (for the definition of
$T'$ see Ref. \protect\cite{Jin}). Experimental values (full symbols) with
error bars are taken from Fig. 3. of Ref. \protect\cite{Jin}. Open
symbols denote the theoretical predictions $\Im \omega_1$ of our model
calculation using the dielectric formalism.
\label{fig1}}
\end{figure}

\begin{figure}
\caption{Excitation frequencies for the $m=0$ and $m=2$ modes as a function
of $T'$. Notations for $T'$ and for the experimental values are the same as
in Fig. \protect\ref{fig1}. Open symbols denote our theoretical values
$\omega_0 + \Re \omega_1$ in units of the axial trap frequency.
\label{fig2}}
\end{figure}

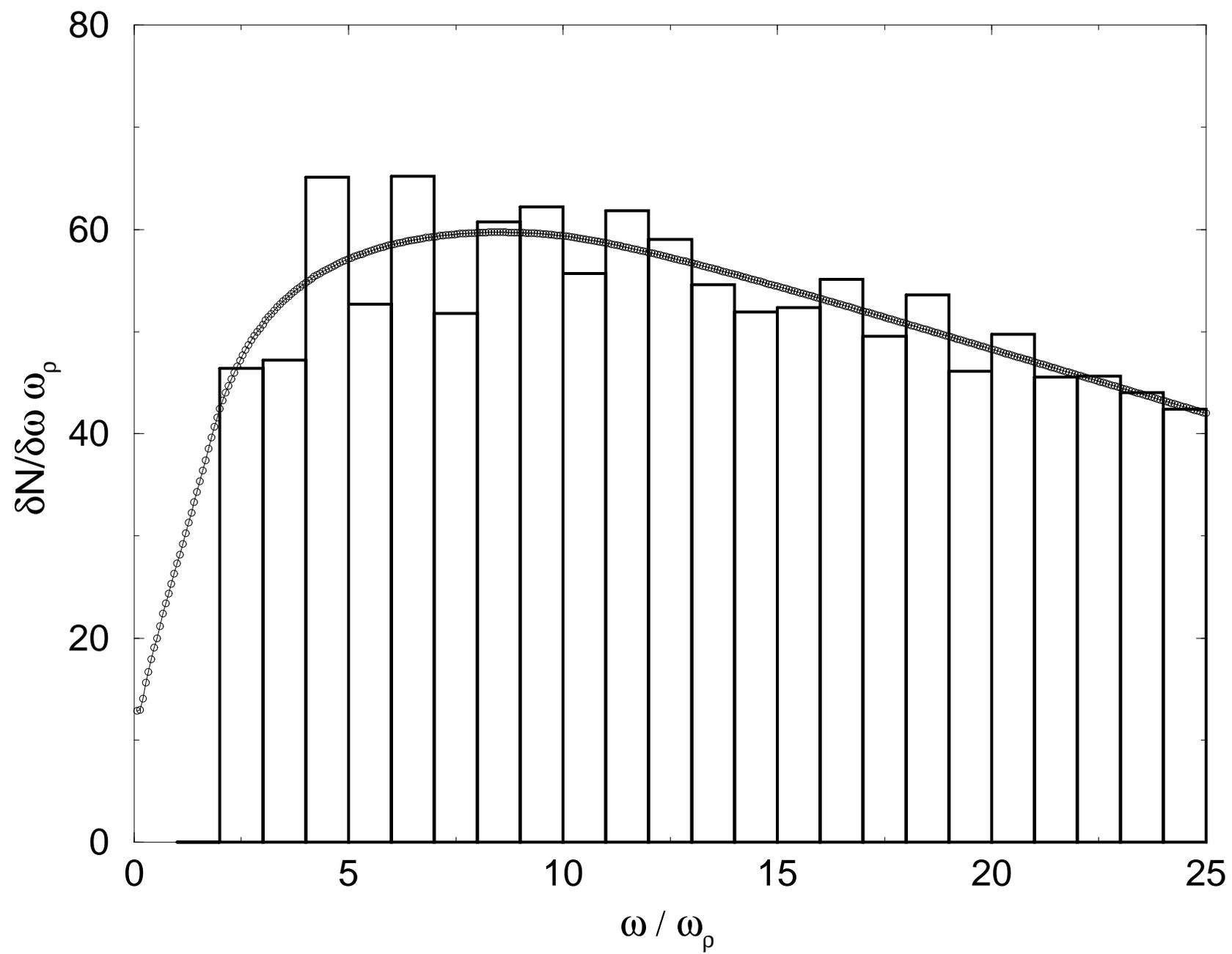
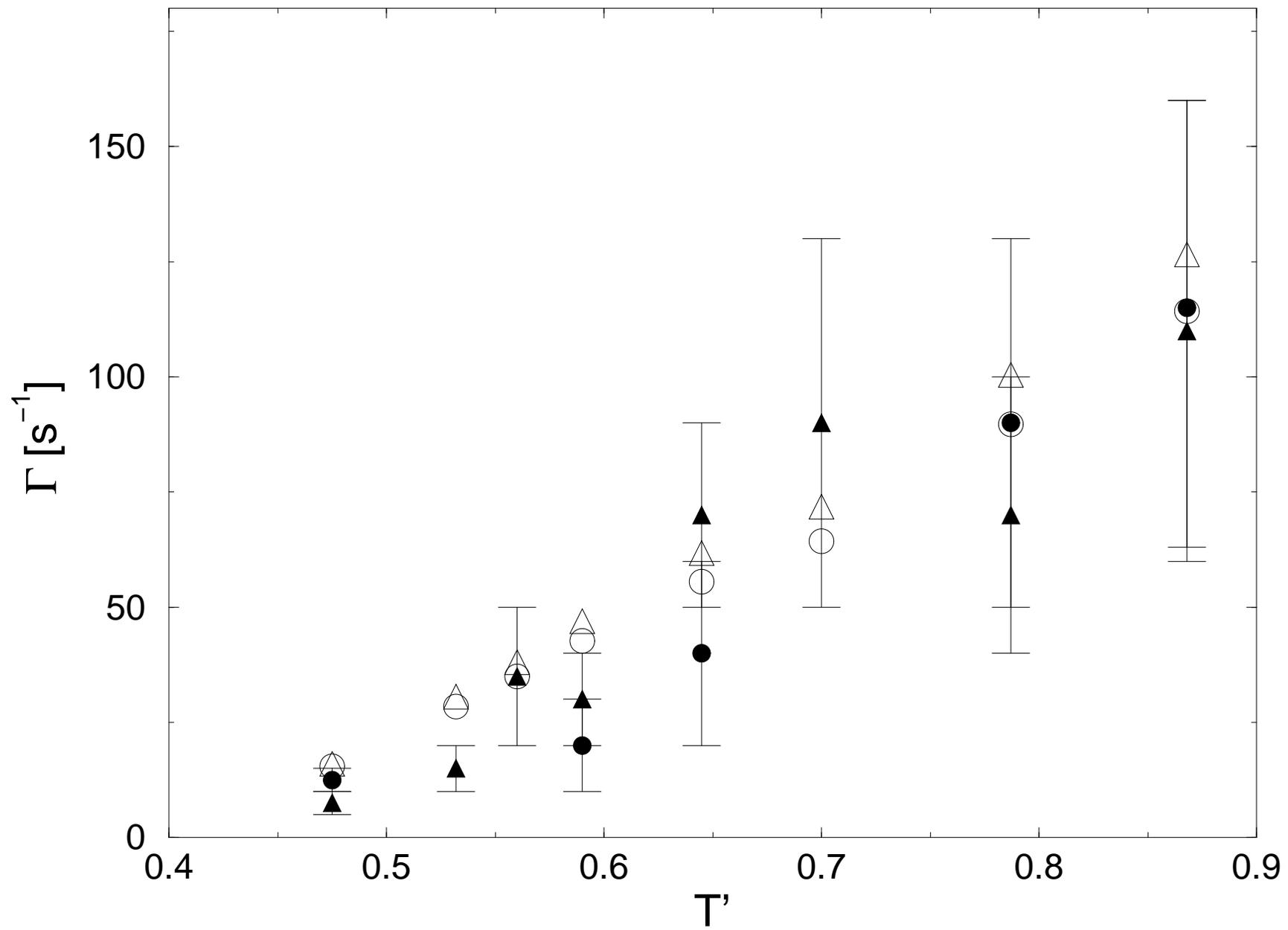
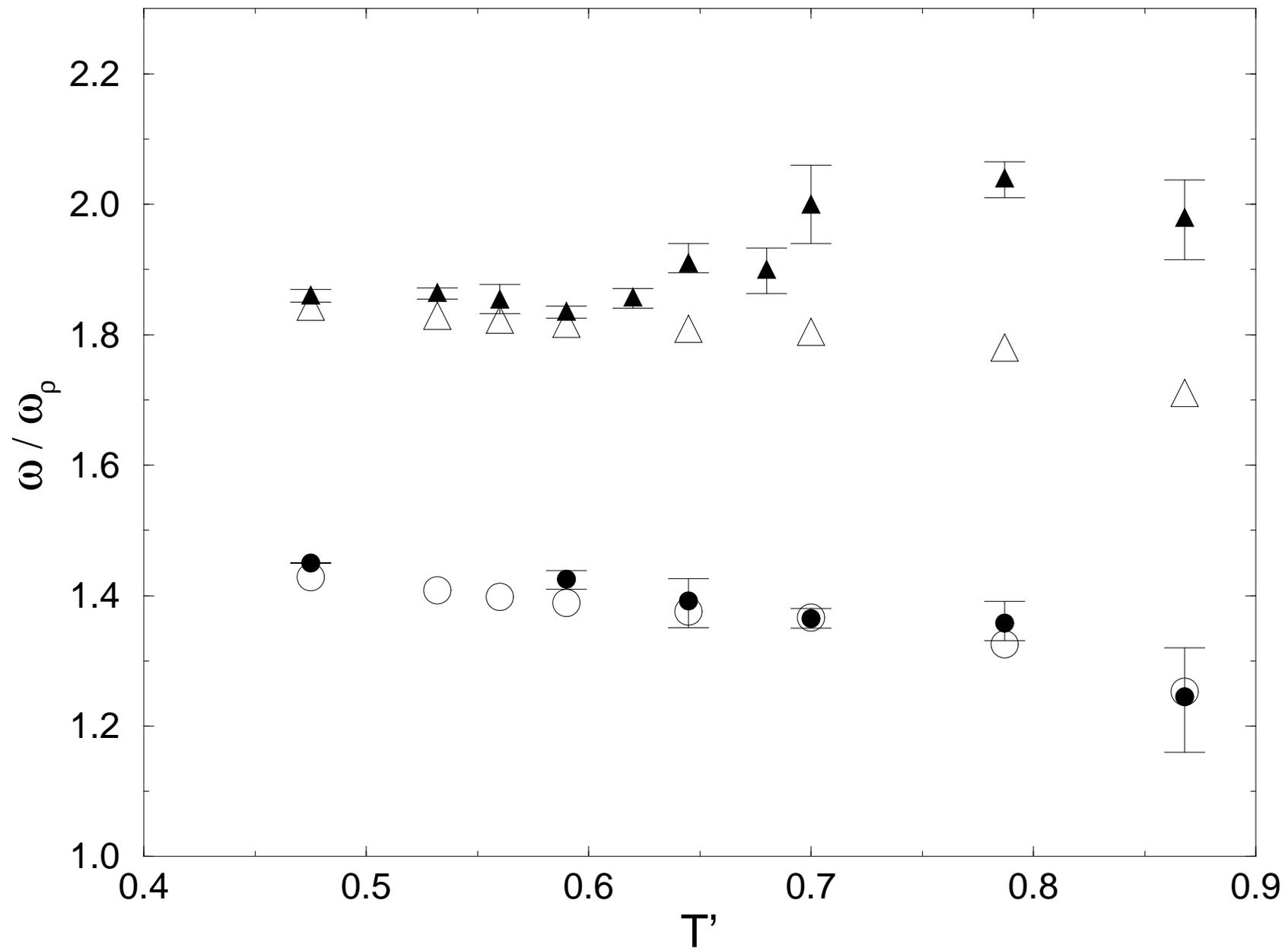
\end{document}